\documentclass[10pt,twocolumn,showpacs,preprintnumbers,amsmath,amssymb,aps,prb,longbibliography,superscriptaddress]{revtex4-2}
\usepackage{amsmath}
\usepackage{graphicx}
\usepackage{wasysym}
\usepackage{amssymb}
\usepackage{amsfonts}
\usepackage{bm}
\usepackage{natbib}
\usepackage{enumerate}
\usepackage{color}
\usepackage{hyperref}
\usepackage{multirow}

\usepackage{times}
\newcommand{\beq}{\begin{equation}}
\newcommand{\eeq}{\end{equation}}
\newcommand{\bea}{\begin{eqnarray}}
\newcommand{\eea}{\end{eqnarray}}
\newcommand{\bes}{\begin{split}}
\newcommand{\ees}{\end{split}}

\usepackage{cleveref}
\crefname{appendix}{App.}{Apps.}
\crefname{equation}{Eq.}{Eqs.}
\crefname{figure}{Fig.}{Figs.}
\crefname{table}{Tab.}{Tabs.}
\crefname{section}{Sec.}{Secs.}

\mathchardef\nss="711B

\def\ket#1{{\left|#1\right\rangle}}

\def\br{\mathbf{r}}

\def\nss{\mathcal{S}}

\def\be{\begin{eqnarray}}
\def\ee{\end{eqnarray}}
\def\beq{\begin{equation}}
\def\eeq{\end{equation}}

\newlength{\myL}

\begin{document}
\title{Twisted coupled wire model for a moir\'e sliding Luttinger liquid}

\affiliation{Department of Physics, Princeton University, Princeton, New Jersey 08544, USA}
\affiliation{Center for Correlated Matter and School of Physics, Zhejiang University, Hangzhou, 310058, China}
\author{Yichen Hu$^{1}$}
\author{Yuanfeng Xu$^{1,2}$}
\author{Biao Lian$^{1}$}

\begin{abstract}

Recent experiments in twisted bilayer WTe$_2$ revealed the existence of anisotropic Luttinger liquid behavior. To generically characterize such anisotropic twisted bilayer systems, we study a model of a twisted bilayer of two-dimensional (2D) arrays of coupled wires, which effectively form an array of coupled moir\'e wires. We solve the model by the transfer matrix method, and identify quasi-1D electron bands in the system at small twist angles. With electron interactions added, we show that the moir\'e wires have an effective Luttinger parameter $g_\text{eff}$ lower than that of the microscopic wires. This leads to a sliding Luttinger liquid (SLL) temperature regime, in which power-law current voltage relations arise. For parameters partly estimated from WTe$_2$, a microscopic interaction $U\sim3$eV yields a temperature regime of SLL similar to that in the WTe$_2$ experiments.

\end{abstract}
\maketitle
One-dimensional (1D) interacting electrons form Luttinger liquids\cite{Tomonaga1950,Haldane1981,Luttinger1963,Giamarchi2003,Wang2020,Jompol2009}, which show non-Fermi liquid physics such as spin-charge separation and power-law voltage-current relations violating the Ohm's law. In highly anisotropic 2D electron systems, it was proposed that interactions can lead to a sliding Luttinger liquid (SLL) phase or regime \cite{Wen1990,Kivelson2000,Sondhi2001,Ashvin2001,Kane2001,Teo2014,Kane2017,Fuji2017,Stern2018,Hansson2019,Tam2020,Tam2021,Ronny2014,Chamon2016,Fuji2019,Meng2015,Patel2016,Teo2023,Giamarchi2000}, which shows quasi-1D physics analogous to the 1D Luttinger liquid\cite{Du2022,Niu2018,pengjie,Yu2023}. Recent studies of moir\'e systems of twisted homobilayer or heterobilayer 2D materials have enabled engineering of a rich variety of 2D electron systems, such as flat bands with strong interactions \cite{Balents2020,Haddadi2020,Li2021,Tang2020,Wu2018,Shimazaki2020,Regan2020,Wang2020,Yankowitz2019,Liu2020,Kim2017,Lu2019,Cao2018,Cao20182,Burg2019,Cao2020,Chen2019,Yu2023,Hsu2023}. Intriguingly, in the twisted bilayer WTe2 which hosts a rectangular moire pattern, transport experiments \cite{pengjie,Yu2023} reveal a phase that exhibits (1) a strong in-plane electronic anisotropy, (2) a power-law scaled conductance in the hard direction and (3) a nonlinear differential resistance that vanishes at zero bias in the easy direction. The phenomena strongly suggests quasi-1D physics in a SLL regime. The tunability of moir\'e systems may allow a thorough exploration of the physics of SLLs. Therefore, it is important to understand theoretically the mechanism of quasi-1D SLLs from twisted bilayer anisotropic systems, and to investigate its dependence on twist angles and interaction strengths.

In this paper, we study a model of a twisted bilayer of 2D arrays of coupled wires, which effectively forms a network model with interwire and interlayer hoppings. The goal of the model is to give a simplified generic description of twisted bilayer anisotropic systems, and investigate the emergence of an array of coupled moir\'e wires showing SLL physics. At small twist angles, the model can be solved from the real space transfer matrix, which directly gives the Fermi surface of moir\'e bands at the Fermi energy. We find that compared to the microscopic coupled wires, the emergent coupled moir\'e wires show much smaller Luttinger parameter $g_\text{eff}$, implying much stronger correlation effects. For parameters partly estimated from WTe$_2$, a microscopic interaction $U\gtrsim 3$eV yields an SLL temperature regime around that in experiments.

\begin{figure}[tbp!]
\centering\includegraphics[width=0.45\textwidth]{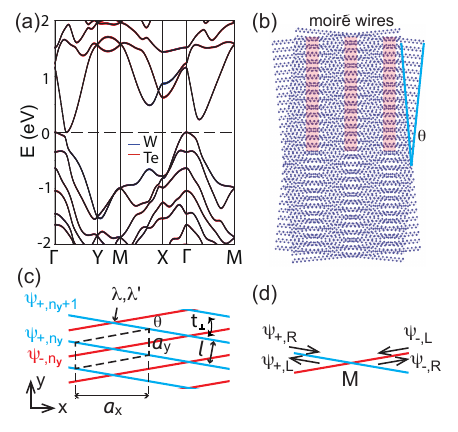}
\caption{(a) DFT band structure of monolayer $WTe_2$, showing domination by $W$ orbitals in the valence band around the $\Gamma$ point. (b) $W$ atom positions in twisted bilayer $WTe_2$, and illustration of the emergent moir\'e wires. (c) Twisted network model consisting of two arrays of coupled wires with a twist angle $\theta$. Dotted lines encircle a moir\'e unit cell. (d) Zoom-in around a single interlayer intersection point, across which the wavefunctions are related by transfer matrix $M$.  }\label{fig:1}
\end{figure}

\indent \emph{The single-particle model.} Monolayer WTe$_2$ has an anisotropic crystal structure and an anisotropic band structure, as shown in Fig. \ref{fig:1}(a). In the valence band which is relevant to SLL experimentally \cite{pengjie,Yu2023}, the $W$ atom orbitals dominate. In real space, the $W$ atoms approximately form arrays of quasi-1D wires [along the horizontal direction in Fig. \ref{fig:1}(b)]. In twisted bilayer WTe$_2$, the $W$ atom quasi-1D wires of the two layers form a moir\'e pattern in Fig. \ref{fig:1}(b), which visually show effective vertical moir\'e wires at a larger length scale.

To give a generic simplified single-particle model of an anisotropic moir\'e system such as twisted bilayer WTe$_2$, we approximate each layer by an array of coupled wires with interwire distance $l$ as shown in \cref{fig:1}(c), where the blue and red wires are in the upper and lower layers, respectively. The two layers differ by a twist angle $\theta$. We assume the electrons (or holes via a particle-hole transformation) in each wire have a quadratic dispersion with an effective mass $m$, and a hopping $-t_\perp$ between two nearest neighbor wires in each layer. Since the relevant hole band in WTe$_2$ is around the $\Gamma$ point, the spin-orbital coupling can be neglected (as is known for transition-metal dichalcogenides \cite{Angeli2021,TMD1}), and thus we can choose $t_\perp$ to be real. As shown in \cref{fig:1}(c), the wires of the two layers cross at positions with $x$ coordinates $x=n_x a_x$, where $n_x\in\mathbb{Z}$, and $a_x=\frac{l}{2\sin(\theta/2)}$. The vertical distance between two neighboring wires in each layer is $a_y=\frac{l}{\cos(\theta/2)}$. At each crossing of two wires from two layers [\cref{fig:1}(d)], we assume a delta function intralayer potential of strength $\lambda$ and a delta function interlayer hopping of strength $\lambda'$. For small twist angle $\theta$, we can approximate the coordinate along each wire by $x$, and we use $\psi_{\alpha,n_y}(x)$ to denote the wavefunction in the $n_y$-th wire ($n_y\in\mathbb{Z}$) in layer $\alpha=\pm$. Note that because of the relative twisting, the $n_y$-th wire in layer $\alpha$ crosses with the $(n_y-\alpha n_x)$-th wire in layer $-\alpha$ at $x=n_xa_x$. The single-particle Shr\"odinger equation at energy $E$ (measured from band bottom) is then
\begin{equation}\label{eq:shrodinger}
\begin{split}
E\psi_{\alpha,n_y}=&\left(2t_\perp-\frac{\hbar^2\partial_x^2}{2m}\right)\psi_{\alpha,n_y}-t_\perp (\psi_{\alpha,n_y+1}+\psi_{\alpha,n_y-1}) \\
&+\delta(x-n_x a_x)\left(\lambda \psi_{\alpha,n_y} +\lambda' \psi_{-\alpha,n_y-\alpha n_x} \right)\ .
\end{split}
\end{equation}

While this oversimplified model does not correspond to a concrete material, our goal is to understand how moir\'e patterns generically give rise to SLL physics. Hereafter, to reflect the physics of twisted bilayer WTe$_2$ as much as possible, we take $l=0.627$nm, interlayer distance $a_z=0.77$nm, and $m=0.38m_e$ where $m_e$ is the bare electron mass, as estimated for the W atom chains in WTe$_2$ from density functional theory (DFT). For the effective interwire hoppings, we take $t_\perp=20$meV, $\lambda=10 \text{ meV} \cdot \text{nm}$ and $\lambda'=20 \text{ meV} \cdot \text{nm}$, based on the order of magnitude of direct hopping between W atoms in DFT [see Appendix \ref{AppC} and \ref{AppD} for details (see also references\cite{vasp1,vasp2,PBE} therein)]. However, we do not expect the above effective parameters to characterize WTe$_2$ quantitatively, which experimentally shows substantial deviations from DFT in some aspects, possibly due to strong interactions\cite{Sun2021,Jia2021,Jing2021,Kwak2018,Liu2015,Crpel2022,Tang2017,Jia2021,Yves2021} 

The Schr\"odinger \cref{eq:shrodinger} for small $\theta$ can be most easily solved by the transfer matrix method. We consider the transfer matrix in the $x$ direction for a state with energy $E$ and quasimomentum $q_y$ in the $y$ direction. For $(n_x-1) a_x<x<n_x a_x$, such a state takes the form
\begin{equation}
\begin{split}
&\psi_{\alpha,n_y}(x)=e^{iq_ya_y(n_y-\frac{\alpha}{2}n_x)}\widetilde{\psi}_{\alpha,q_y}^{n_x}(x)\ ,\\
&\widetilde{\psi}_{\alpha,q_y}^{n_x}(x)=e^{ik(x-n_xa_x)}\varphi_{\alpha,q_y}^{n_x,R}+e^{-ik(x-n_xa_x)}\varphi_{\alpha,q_y}^{n_x,L}\ ,
\end{split}
\end{equation}
where $k=\hbar^{-1}\sqrt{2m[E+2t_\perp (\cos(q_y a_y)-1)]}$ is the free momentum along the wire, and $\varphi_{\alpha,q_y}^{n_x,R}$ and $\varphi_{\alpha,q_y}^{n_x,L}$ are the right-moving and left-moving amplitudes, respectively. By re-organizing the wavefunction into a four-component vector $\Psi_{q_y,n_x}=(\varphi_{+,q_y}^{n_x,L},\varphi_{+,q_y}^{n_x,R},\varphi_{-,q_y}^{n_x,L},\varphi_{-,q_y}^{n_x,R})^T$, the Shr\"odinger equation can be reformulated into the transfer matrix equation (Appendix \ref{AppA})
\begin{equation}
\Psi_{q_y,n_x+1}=T(E,q_y)\Psi_{q_y,n_x}\ .
\end{equation}
The transfer matrix can be further decomposed into two parts, $T(E,q_y)=Q(E,q_y)M(E,q_y)$, where $Q(E,q_y)=\text{diag}(e^{-ika_x},e^{ika_x},e^{-ika_x},e^{ika_x})$ is the diagonal propagation matrix, and one can show that
\begin{equation}
M(E,q_y)=\begin{pmatrix}
(\sigma_0-\frac{i \lambda' m}{k}s_0)e^{i \frac{q_y}{2}} & -\frac{i \lambda m}{k}s_0e^{i \frac{q_y}{2}} \\
-\frac{i \lambda m}{k}s_0 e^{-i \frac{q_y}{2}} & (\sigma_0-\frac{i \lambda' m}{k}s_0)e^{-i \frac{q_y}{2}}\\
\label{Eq4}
\end{pmatrix}
\end{equation}
is the scattering matrix at the node, where $\sigma_0$ is the $2\times2$ identity matrix, and $s_0=\sigma_z+i\sigma_y$ in terms of the $2\times2$ Pauli matrices $\sigma_{x,y,z}$.

One can solve the eigenvectors of the transfer matrix $T(E,q_y)$, and we denote the eigenvalue as $e^{iq_xa_x}$:
\begin{equation}
T(E,q_y)\widetilde{\Psi}_{q_y,q_x}=e^{iq_xa_x}\widetilde{\Psi}_{q_y,q_x}\ .
\end{equation}

If $q_x$ is real, the eigenvector $\widetilde{\Psi}_{q_y,q_x}$ gives an eigenstate of the Shr\"odinger \cref{eq:shrodinger}, with quasimomentum $q_x$ in the $x$ direction and $q_y$ in the $y$ direction.

\begin{figure}[tbp]
\centering\includegraphics[width=0.5\textwidth]{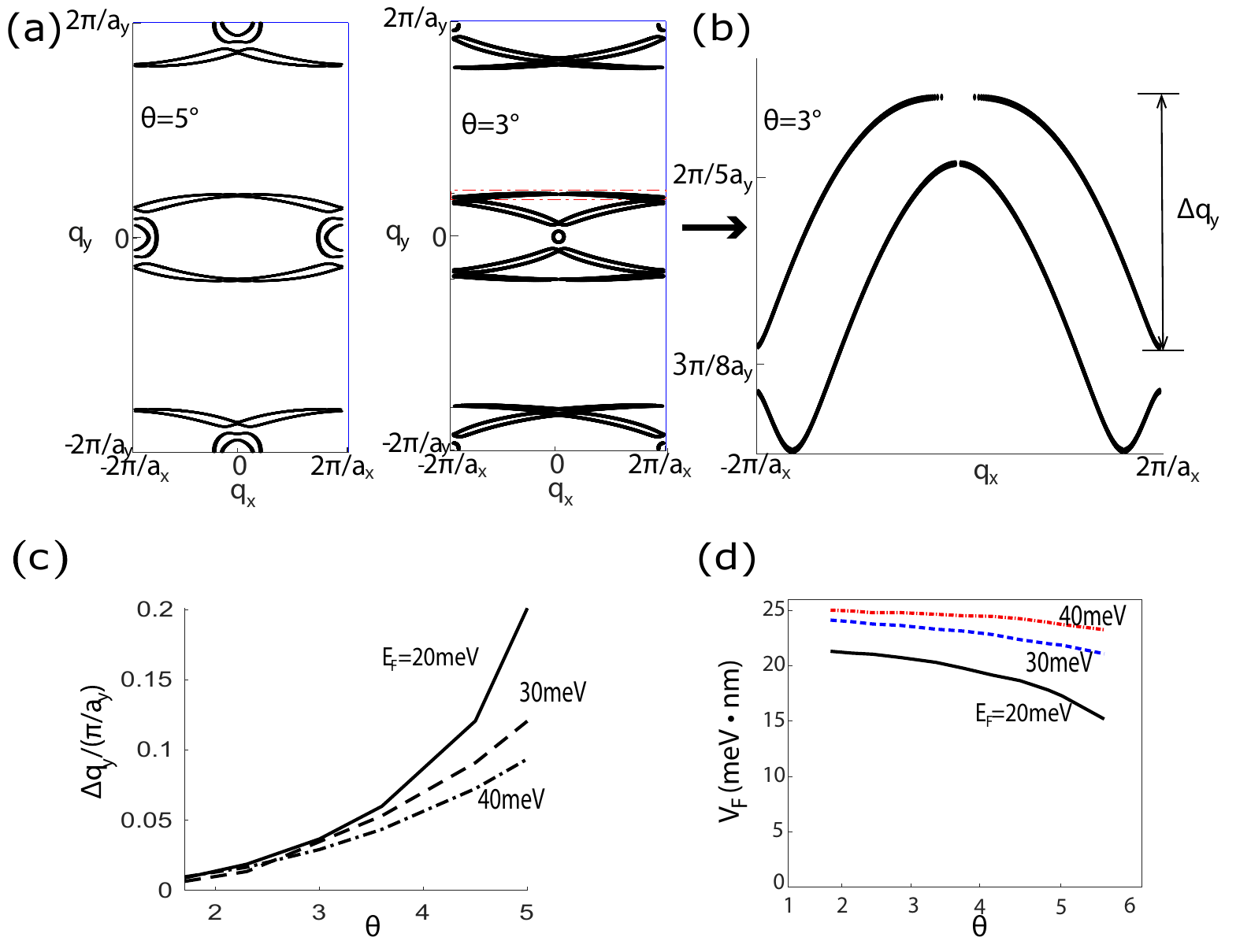}
\caption{(a) Fermi surface solved for $\theta=5^\circ$ (left) and $\theta=3^\circ$ (right) twisted model with Fermi energy $E_F=30$meV, using transfer matrix method (with parameters given below \cref{eq:shrodinger}). (b) Zoomed-in plot of red dotted region of the $3^\circ$ Fermi surface in (a). $\Delta q_y$ characterizes the flatness of $q_x$-direction dispersion of the band. (c) $\frac{\Delta q_y}{\pi/a_y}$ versus $\theta$ for different $E_F$. (d) Fermi velocity $v_F$ versus angle $\theta$ for different Fermi energy $E_F$.}\label{fig:2}
\end{figure}

Taking $E=E_F$ to be the Fermi energy, the above procedure naturally gives the Fermi surface of the band structure in the quasimomentum $(q_x,q_y)$ space. The transfer matrix is thus advantageous if one focuses on the low energy states near the Fermi surface. \cref{fig:2}(a) shows the Fermi surface (black lines) calculated for $E_F=30$meV at $5^\circ$ (left) and $3^\circ$ (right). \cref{fig:2}(b) shows zoomed in image of bands boxed in red dotted lines in $3^\circ$ Fermi surface. As shown in \cref{fig:2}(a), there are Fermi surfaces forming a periodic loop in the $q_x$ direction, indicating that these bands are highly quasi-1D, with effective coupled moir\'e wires as shown in \cref{fig:1}(b). We focus on the band with the flattest Fermi surface. We characterize the anisotropy of the band by the dimensionless number $a_y\Delta q_y/\pi$, where $\Delta q_y$ is the width of $q_y$ spanned by the Fermi surface dispersion shown in \cref{fig:2}(b). This anisotropy as a function of twist angle $\theta$ for different $E_F$ is shown by the black lines in \cref{fig:2}(c). We note that such quasi-1D Fermi surfaces forming loops in $q_x$ only occur below a certain twist angle $\theta_c$, for instance, $\theta_c\approx 6.3^\circ$ for $E_F=30$meV. Moreover, we can numerically calculate the Fermi velocity of the band $v_F=dE/dq_y$ at $q_x=0$ by slightly varying $E$ and $q_y$. The Fermi velocity $v_F$ as a function of $\theta$ for different $E_F$ is given in \cref{fig:2}(d).

We can estimate the effective $y$ direction hopping energy of a moir\'e wire in \cref{fig:1}(b) by $\widetilde{t}_\parallel=\pi v_F/a_y$, using the fact that $\pi/a_y$ is the half Brillouin zone size in the $y$ direction. The effective hopping across two nearest moir\'e wires can be estimated by $\widetilde{t}_\perp=v_F\Delta q_y$, which is the energy scale inducing the Fermi surface dispersion of width $\Delta q_y$. Therefore, the anisotropy of effective coupled moir\'e wires in \cref{fig:2}(c) is also equal to
\begin{equation}
\widetilde{t}_\perp/\widetilde{t}_\parallel=a_y\Delta q_y/\pi\ .
\end{equation}
The smaller $\widetilde{t}_\perp/\widetilde{t}_\parallel$ is, the more quasi-1D in $y$ direction the moir\'e band is.

\emph{Interaction effect.} We now add electron-electron interaction to the model, and examine the emerging SLL physics in the moir\'e scale. The emergent quasi-1D moir\'e bands in the above model provide fertile ground for interaction induced correlations. In the microscopic lattice scale, we consider a generic \emph{spin-independent} anisotropic screened repulsive interaction between two electrons with distance $\br$ (both intralayer and interlayer):
\begin{equation}
V_I(\br)=U \exp\left(-\sqrt{\frac{x^2}{x_0^2}+\frac{y^2}{y_0^2}+\frac{z^2}{z_0^2}}\right)\ .
\label{V_I}
\end{equation}
In a 2D Thomas-Fermi screening model of Coulomb interaction, the screening length is estimated as $r_0=\frac{2\pi\epsilon_r\epsilon_0\hbar^2}{\text{e}^2 m}$ along a direction with effective mass $m$, where $\text{e}$ is the electric charge. Along the $x$-~direction, taking $\epsilon_r\approx 10$ (which is approximately the value for WTe$_2$\cite{Domozhirova2019}), and effective mass $m=0.38m_e$, one arrives at a screening length $x_0\sim 1$nm. The quasi-1D nature (along the $x$ direction) of the microscopic system gives heavier effective masses (by approximately one order of magnitude) along the $y$ and $z$ directions, so we take $y_0=5$nm, $z_0=10$nm. We note that for low carrier densities  with which we are concerned, the bare interaction energy scale is likely larger than the electron kinetic energy, so the Thomas-Fermi model may overestimate the screening. Nevertheless, as a model study, we fix $x_0$, $y_0$, and $z_0$ as above, and leave the microscopic interaction $U$ as a tuning parameter. For reference, taking $V_I(x_0,0,0)\sim\frac{\text{e}^2}{4\pi\epsilon_0 x_0}$ yields a $U$ of order $5$eV.

For a 1D moir\'e wire, the electrons with interaction becomes a Luttinger liquid with spin-charge separation. The charge Luttinger parameter $g_{\text{eff}}$ and spin Luttinger parameter $g_{\text{eff},s}$ are given by \cite{Fisher1997,Giamarchi2003}
\begin{equation}\label{eq:geff}
\begin{split}
&g_{\text{eff}}=\sqrt{\frac{2\pi v_F+\widetilde{V}(2q_F)}{2\pi v_F+2\widetilde{V}(0)-\widetilde{V}(2q_F)}}\ ,\\
&g_{\text{eff},s}=\sqrt{\frac{2\pi v_F+\widetilde{V}(2q_F)}{2\pi v_F-\widetilde{V}(2q_F)}}\ ,
\end{split}
\end{equation}
where $\widetilde{V}(0)$ and $\widetilde{V}(2q_F)$ are the interaction strength in the quasimomentum $q_y$ space for scattering with zero momentum change (forward scattering) and $2q_F$ momentum change (back scattering), respectively (see Appendix \ref{AppB}), and $v_F=dE/dq_y$ is the Fermi velocity. For repulsive interaction, one usually has $\widetilde{V}(0)>\widetilde{V}(2q_F)$ and $\widetilde{V}(0)>0$, thus $g_{\text{eff}}<1$. For the parameters we choose, we find $|\widetilde{V}(2q_F)|\sim 10^{-3}\widetilde{V}(0)\ll v_F$, so the spin Luttinger parameter is always approximately $g_{\text{eff},s}\approx 1$. 

\begin{figure}[tbp!]
\centering\includegraphics[width=0.45\textwidth]{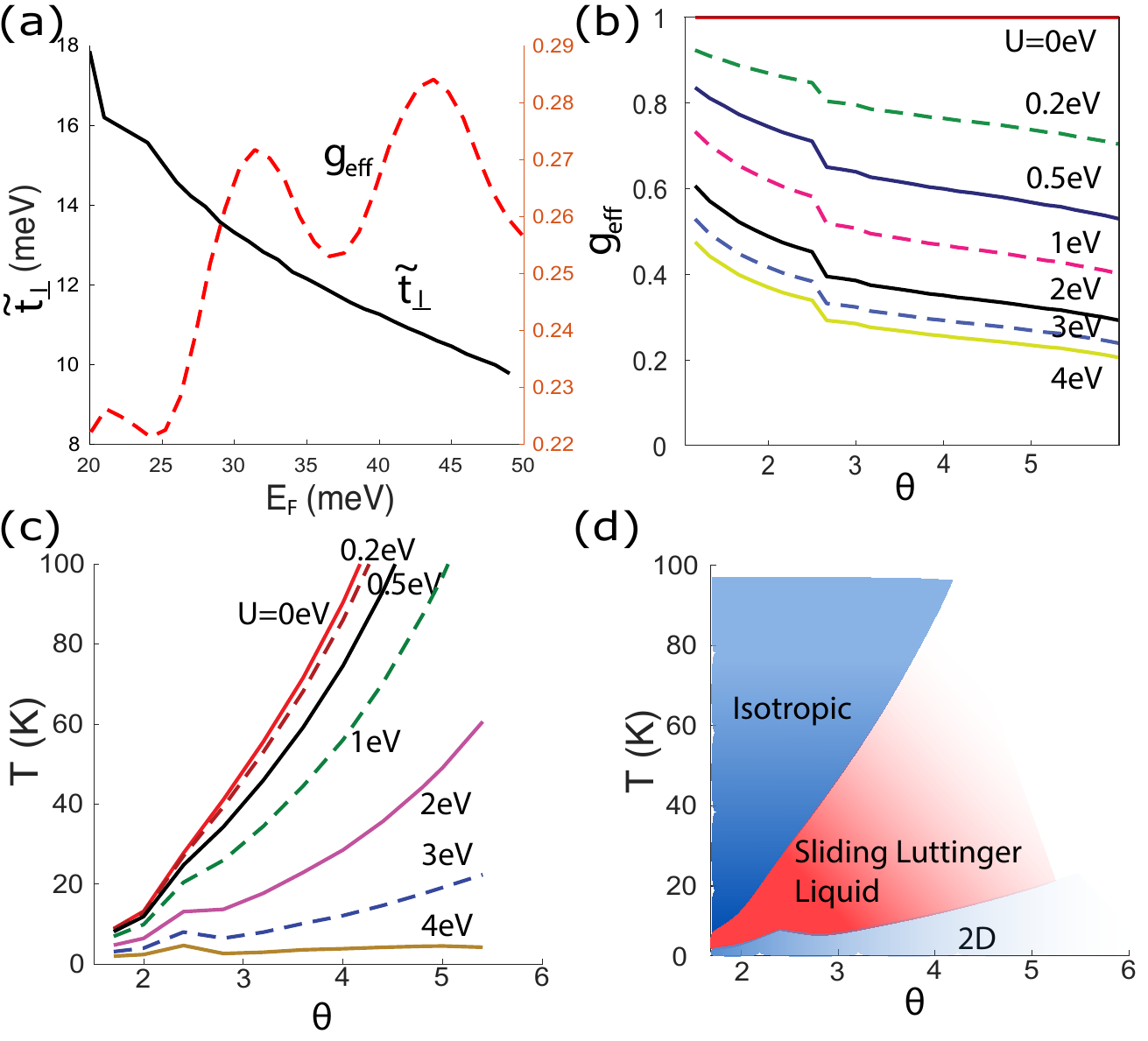}
\caption{(a) Plots of effective Luttinger parameter $g_{\text{eff}}$ and interwire coupling $\widetilde{t}_{\perp}$ vs Fermi energy $E_F$ at $5^\circ$ and $U=3 eV$. The left $y$ axis shows $\widetilde{t}_\perp$ and the right $y$ axis shows $g_{\text{eff}}$. (b) Plot of $g_{\text{eff}}$ versus twist angle $\theta$ for different $U$ at Fermi energy $E_F=30$meV. (c) The lower-bound SLL temperature $T_*$ vs $\theta$ for different $U$ at Fermi energy $E_F=30$meV. (d) Phase diagram for a twisted system at different angles $\theta$ for $U=3$eV at Fermi energy $E_F=30$meV. The upper and lower boundaries are crossover temperatures $T_0$ and $T_*$, respectively. }\label{fig:3}
\end{figure}

We average over pairs of Fermi surface states at quasimomenta $(q_x,\pm q_F(q_x))$ to obtain the average charge Luttinger parameter $g_{\text{eff}}$. \cref{fig:3}(a) shows $g_{\text{eff}}$ as a function of $E_F$ calculated for $U=3$eV and twist angle $\theta=5^\circ$, which is below $0.3$. The effective hopping $\widetilde{t}_\perp$ between neighboring moir\'e wires is also plotted. \cref{fig:3}(b) shows how $g_{\text{eff}}$ depends on the interaction strength $U$ as a function of $\theta$, where $E_F=30$meV. We find that $g_{\text{eff}}$ increases as $\theta$ decreases. This is because the width $a_x$ of a moir\'e wire increases as $\theta$ decreases, and thus two electrons in the moir\'e wire have less chance to scatter with each other, leading to a weaker effective interaction. In comparison, for a microscopic decoupled wire in a monolayer, which has a Fermi momentum $k_F=\sqrt{2m E_F}$, we can estimate its microscopic Luttinger parameter $g$ similarly to \cref{eq:geff}, which is generically larger ($g\gtrsim 0.6$) than the moir\'e effective Luttinger parameter $g_{\text{eff}}$ here given the same $U$. 

\emph{The moir\'e SLL regime.} The temperature regime for SLL physics depends on the effective moir\'e parameters $\widetilde{t}_\perp$, $\widetilde{t}_\parallel$, and $g_{\text{eff}}$\cite{Kane2001,Ashvin2001,Giamarchi2000,Bourbonnais1984}, which we explain below. Without the interwire hopping $\widetilde{t}_\perp$, and setting $g_{\text{eff},s}=1$, each moir\'e wire along the $y$ direction has a zero-temperature 1D Luttinger liquid Green's function following\cite{Fradkin2013}
\begin{equation}\label{eq:G1D}
\begin{aligned}
&G_\zeta^{\text{1D}}(y,t)=\langle \mathcal{T} \hat{\psi}_{n_x,\zeta}(y,t) \hat{\psi}_{n_x,\zeta}^\dag (0,0) \rangle\\
&=\frac{1}{2\pi}(y-\zeta v_c t+i\zeta0^+)^{-\frac{1}{2}}(y-\zeta v_s t+i\zeta0^+)^{-\frac{1}{2}}\\
&\times (y^2-v_c^2(t-i0^+)^2)^{-\frac{\eta}{2}}\ ,
\end{aligned}
\end{equation}
where $\zeta=\pm$ stands for the modes moving in the $\pm y$ directions, respectively, the charge mode has velocity $v_c=\frac{1}{2\pi}\sqrt{[2\pi v_F+\widetilde{V}(0)]^2-[\widetilde{V}(2q_F)-\widetilde{V}(0)]^2}$, the spin mode has velocity $v_s=\frac{1}{2\pi}\sqrt{4\pi^2 v_F^2-\widetilde{V}(2q_F)^2}\approx v_F$, 

and the anomalous exponent $\eta=(g_{\text{eff}}+g_{\text{eff}}^{-1}-2)/4$. In particular, $\eta=0$ in the non-interacting $g_{\text{eff}}=1$ limit. At low energies, the local density of states (spectral weight) scales as
\begin{equation}
\rho(\omega)=-\frac{1}{\pi}\text{Im}\sum_{\zeta=\pm} \widetilde{G}^{1D}_\zeta(0,\omega)\propto \omega^{\eta}\ ,
\end{equation} 
where $\widetilde{G}^{1D}_\zeta(y,\omega)$ is the Fourier transform of \cref{eq:G1D} in time $t$. As can be seen by a simple power counting, the full Fourier transform of \cref{eq:G1D} in both $y$ and $t$ scales as $G_\zeta^{\text{1D}}(q,\omega)= \omega^{\eta-1}\widetilde{t}_\parallel^{-\eta}f(q/\omega)$ (which has the unit of inverse of energy), where $f(x)$ is some dimensionless function of order unity.

With the effective interwire hopping $\widetilde{t}_\perp$, the hopping among different moir\'e wires can be incorporated as a self energy correction to \cref{eq:G1D}. The zero-temperature Green's function is perturbatively given by a Schwinger-Dyson equation \cite{Bourbonnais1984}:
\begin{equation}\label{eq:G2D}
\begin{aligned}
G_\zeta(q,\omega)&\approx G^{1D}_\zeta(q,\omega)+\widetilde{t}_\perp G^{1D}_\zeta(q,\omega) G_\zeta(q,\omega)\\
&\approx\frac{G^{1D}_\zeta(q,\omega)}{1-\widetilde{t}_\perp G^{1D}_\zeta(q,\omega)}\ .
\end{aligned}
\end{equation}
In the case $0\le\eta<1$, for which the interaction is not too strong, $G_\zeta^{\text{1D}}(q,\omega)\propto \omega^{\eta-1}\widetilde{t}_\parallel^{-\eta}$ diverges in the low energy limit $\omega\rightarrow 0$. Clearly, the perturbation estimation \cref{eq:G2D} will break down when $\widetilde{t}_\perp G^{1D}_\zeta(q,\omega)\sim \widetilde{t}_\perp \widetilde{t}_\parallel^{-\eta} \omega^{\eta-1} > 1$, namely, when the energy scale $\omega< \widetilde{t}_\perp (\widetilde{t}_\perp/\widetilde{t}_\parallel)^{\frac{\eta}{1-\eta}}$. In this case, $\widetilde{t}_\perp$ can no longer be treated as a perturbation, and the system will behave as an intrinsically 2D system. For energy scales $\omega> \widetilde{t}_\perp (\widetilde{t}_\perp/\widetilde{t}_\parallel)^{\frac{\eta}{1-\eta}}$, the Green's function in \cref{eq:G2D} resembles that of the 1D Luttinger liquid closely, and SLL quasi-1D behaviors are expected. In the case $\eta\ge 1$, $G_\zeta^{\text{1D}}(q,\omega)$ does not diverge as $\omega\rightarrow 0$, and thus the SLL quasi-1D behavior persists down to $\omega=0$.

At finite temperature, the temperature $T$ plays the role of the energy scale of the system, and thus the 1D Luttinger liquid Green's function $G^{1D}_\zeta(q,\omega)\propto (k_BT)^{\eta-1}\widetilde{t}_\parallel^{-\eta}$ in the low energy limit $\omega\rightarrow 0$, where $k_B$ is the Boltzmann constant. Therefore, the quasi-1D SLL physics can only exist for temperatures $T>T_*$, where the lower bound crossover temperature for $0\le \eta<1$ is
\begin{equation}\label{eq:lowT}
T_*\approx k_B^{-1}\widetilde{t}_\perp \left(\widetilde{t}_\perp/\widetilde{t}_\parallel\right)^{\frac{\eta}{1-\eta}}
\end{equation}
For $\eta\ge1$, the SLL physics will persist down to $T_*=0$.

In the SLL temperature regime $T>T_*$, the transverse conductivity $\sigma_\perp$ across the moir\'e wires [namely, in the $x$ direction in \cref{fig:1}(c)] at temperature $T$ is known to exhibit a power law scaling \cite{Wen1990,Kivelson2000,Sondhi2001,Ashvin2001,Kane2001,Teo2014,Giamarchi2000}, Treating $\widetilde{t}_\perp$ as perturbation, $\sigma_\perp$ is governed by the Kubo formula for tunneling between two neighboring wires:
\begin{equation}\label{eq:kubo}
\begin{split}
\sigma_\perp(\omega,T)\propto\ & \widetilde{t}_\perp^2\int dy\int d\omega'\frac{n_f(\omega'+\omega)-n_f(\omega')}{\omega} \\
&\quad \times \text{Im}\widetilde{G}^{1D}_\alpha(y,\omega')\text{Im}\widetilde{G}^{1D}_\alpha(y,\omega+\omega') \\
\propto \ & \widetilde{t}_\perp^2 T^{2\eta -1}F\left(\frac{\omega}{T}\right)\ ,
\end{split}
\end{equation}
where $F(x)$ approaches $1$ as $x\rightarrow 0$, and scales as $x^{2\eta-1}$ as $x\gg 1$. For transverse transport measurement with voltage $V$ between two neighboring moir\'e wires, the energy scale $\omega$ can be identified with $V$. Therefore, the transverse conductivity
\begin{equation}\label{eq:nonlinear}
\begin{split}
\sigma_\perp(V,T)\propto 
\begin{cases}
V^{2\eta-1}\ ,\quad  (eV>k_BT)\\
T^{2\eta-1}\ ,\quad  (eV<k_BT) 
\end{cases}
\end{split}
\end{equation}
which is the key feature of SLL. The longitudinal conductivity $\sigma_\parallel$ along the moir\'e wires [the $y$ direction in \cref{fig:1}(c)] is expected to be similar to that of the 1D Luttinger liquid, which increases as $T$ decreases in the disorderless limit (diverging at $T=0$, if temperature $T$ could reach zero).

Physically, the SLL regime cannot persist to arbitrarily high temperature, since the Luttinger liquid theory is only valid at low energies. In the literature \cite{Giamarchi2003,Biermann2002}, it is assumed that the upper bound temperature is $T_0\sim \widetilde{t}_\parallel/k_B$, which is a very high temperature. Here we argue that the upper bound temperature for the power law behavior of $\sigma_\perp(V,T)$ in \cref{eq:nonlinear} is much lower, due to the presence of marginal and irrelevant couplings (inelastic scattering with phonons, etc.). Note that $\sigma_\perp(V,T)$ in \cref{eq:kubo} considers only the relevant bare coupling $\widetilde{t}_\perp$ which has the unit of energy. In general, irrelevant couplings of dimension $-\Delta<0$ would induce effective couplings scaling as $(k_BT)^\Delta$, which grows as $T$ increases. Without a detailed knowledge of the irrelevant couplings, we expect their corrections to be no longer negligible when $k_BT>\widetilde{t}_\perp$. This sets an upper bound crossover temperature for the SLL behavior as 
\begin{equation}\label{eq:highT}
T_0\sim k_B^{-1}\widetilde{t}_\perp\ , 
\end{equation}
which we adopt here. The SLL regime is thus within the temperature range $T_*<T<T_0$.

\cref{fig:3}(c) shows the lower bound temperature $T_*$ as a function of twist angle $\theta$ for $E_F=30$meV and interaction strength $U$ ranging from $0$ to $4$eV, for which the Luttinger liquid parameter $g_{\text{eff}}$ is shown in \cref{fig:2}(b). Generically, we find that $T_*$ decreases as $\theta$ decreases, because of the decrease of $\widetilde{t}_\perp$, although $g_{\text{eff}}$ increases. As an example, \cref{fig:3}(d) shows the phase diagram plotted for $E_F=30$meV , $U=3$eV with respect to $\theta$, where the lower and upper curves are the crossover temperatures $T_*$ in \cref{eq:lowT} and $T_0$ and \cref{eq:highT}, respectively, in between which is the SLL regime. We note that $T_*$ and $T_0$ are both order estimations. At temperatures $T>T_0$, one expects the anisotropy of the system to be suppressed by strong thermal fluctuations and irrelevant couplings, which we call the isotropic regime. For temperatures $T<T_*$ which we denote as the 2D regime, the perpendicular hopping $\widetilde{t}_\perp$ becomes important, and the system will behave intrinsically as a low temperature 2D system. As the temperature $T$ decreases, depending on interactions, the system may remain a 2D Fermi liquid till $T=0$, or undergo a symmetry breaking phase transition into charge density wave, etc., as studied in the literature \cite{Kivelson2000,Sondhi2001,Ashvin2001,Kane2001}. 

\emph{Discussion}. We have seen that for the effective coupled moir\'e wires, the Luttinger parameter $g_{\text{eff}}$ (e.g., $\sim0.3$ for $U=3$eV) is much lower than the Luttinger parameter $g$ ($\sim 0.6$ for $U=3$eV, see Appendix \ref{AppC}) of the monolayer microscopic coupled wires. Generically, this is because of the lowering of Fermi velocity $v_F$ and effectively longer range of interaction broadened by the widths of moir\'e wires. The lower bound temperature $T_*$ of the SLL regime can be rather sensitive to $U$. For our parameters, at $\theta=5^\circ$, $T_*$ drops from $50$K to about $4$K as $U$ increases from $2$ to $4$eV. If $U$ further increases, $T_*$ would drop down to zero, leading to a zero-temperature SLL quantum phase. In the WTe$_2$ experiments \cite{pengjie,Yu2023}, $T_*\sim 2$K at $\theta=5^\circ$, and $T_*<50$mK at $\theta=3^\circ$, which are similar to our results for $U\gtrsim 3$eV. This suggests that a strong microscopic interaction is needed to explain the experiments. Future studies with more accurate moir\'e models are needed for quantitative understandings. More generally, we expect our twisted coupled wire modeling to be applicable to twisted bilayers of 2D anisotropic systems, such as arrays of nanowires \cite{Jompol2009,Wang2020}, black phosphorus \cite{wangh2023}, etc., with different energy scales.

\begin{acknowledgments}
Acknowledgments. We are thankful for helpful discussions with Tiancheng Song, Guo Yu, Pengjie Wang, Sanfeng Wu, and Jing Wang. This work is supported by the Alfred
P. Sloan Foundation, by the National Science Foundation
under award (DMR-2141966) and (DMR-2011750) through
the Princeton University’s Materials Research Science and
Engineering Center. Additional support was provided by the
Gordon and Betty Moore Foundation through Grant No.
GBMF8685 towards the Princeton theory program.
\end{acknowledgments}

\newpage

\appendix
\onecolumngrid

\section{Transfer Matrix}\label{AppA}
In this section, we supplement details on how to obtain transfer matrix that helped us solving band structures for our bilayer system.

Since we have a network model of quantum wires, intersection points form a lattice $\mathbb{Z}^2$. In real space, we can relate wavefunctions passing through a column of intersection points along $y$-direction using transfer matrix $M$:
\begin{equation}
\begin{split}
\centering &\begin{pmatrix}
\vdots\\
\psi_{+,n_y}^L(n_xa_x+0^+)\\
\psi_{+,n_y}^R(n_xa_x+0^+)\\
\psi_{-,n_y-n_x}^L(n_xa_x+0^+)\\
\psi_{-,n_y-n_x}^R(n_xa_x+0^+)\\
\psi_{+,n_y+1}^L(n_xa_x+0^+)\\
\psi_{+,n_y+1}^R(n_xa_x+0^+)\\
\psi_{-,n_y-n_x+1}^L(n_xa_x+0^+)\\
\psi_{-,n_y-n_x+1}^R(n_xa_x+0^+)\\
\vdots
\end{pmatrix}\\
&=\begin{pmatrix}
\ddots&&&&&&&\\
0&0&0&0&1-\frac{i \lambda' m}{k}&-\frac{i \lambda' m}{k}&-\frac{i \lambda m}{k}&-\frac{i \lambda m}{k}\\
0&0&0&0&\frac{i \lambda' m}{k}&1+\frac{i \lambda' m}{k}&\frac{i \lambda m}{k}&\frac{i \lambda m}{k}\\
-\frac{i \lambda m}{k}&-\frac{i \lambda m}{k}&1-\frac{i \lambda' m}{k}&-\frac{i \lambda' m}{k}&0&0&0&0\\
\frac{i \lambda m}{k}&\frac{i \lambda m}{k}&\frac{i \lambda' m}{k}&1+\frac{i \lambda' m}{k}&0&0&0&0\\
1-\frac{i \lambda' m}{k}&-\frac{i \lambda' m}{k}&-\frac{i \lambda m}{k}&-\frac{i \lambda m}{k}&0&0&0&0\\
\frac{i \lambda' m}{k}&1+\frac{i \lambda' m}{k}&\frac{i \lambda m}{k}&\frac{i \lambda m}{k}&0&0&0&0\\
0&0&0&0&-\frac{i \lambda m}{k}&-\frac{i \lambda m}{k}&1-\frac{i \lambda' m}{k}&-\frac{i \lambda' m}{k}\\
0&0&0&0&\frac{i \lambda m}{k}&\frac{i \lambda m}{k}&\frac{i \lambda' m}{k}&1+\frac{i \lambda' m}{k}\\
&&&&&&&\ddots
\end{pmatrix}
\begin{pmatrix}
\vdots\\
\psi_{+,n_y}^L(n_xa_x-0^+)\\
\psi_{+,n_y}^R(n_xa_x-0^+)\\
\psi_{-,n_y-n_x}^L(n_xa_x-0^+)\\
\psi_{-,n_y-n_x}^R(n_xa_x-0^+)\\
\psi_{+,n_y+1}^L(n_xa_x-0^+)\\
\psi_{+,n_y+1}^R(n_xa_x-0^+)\\
\psi_{-,n_y-n_x+1}^L(n_xa_x-0^+)\\
\psi_{-,n_y-n_x+1}^R(n_xa_x-0^+)\\
\vdots.
\end{pmatrix}
\end{split}
\end{equation}

If we perform discrete Fourier transformation on the above transfer matrix $M$ along $y$-direction, we obtain Eq.~(4).

\section{Inter and Intra-layer interactions}\label{AppB}
In this section, we present details on our calculations of $k=0$ and $2q_y$ component of potential $V(\mathbf{r})$.

Moir\'e wavefunctions within a unit cell at momentum $q_x,q_y$ is 
\begin{equation}
\psi_{\alpha,q_y,q_x}(\mathbf{r})=(\psi_{\alpha}^L e^{-ikx}+\psi_{\alpha}^R e^{ikx})
\end{equation} where $\mathbf{r}=(x,-\alpha\frac{a_y}{a_x}x,\alpha \frac{a_z}{2})$ and $a_z=0.77$nm is the interlayer distance.
We require wavefunction be normalized in a unit cell:
\begin{equation}
\sum_{\alpha=\pm}\int_{uc} d\mathbf{r} |\psi_{\alpha,q_y,q_x}(\mathbf{r})|^2=1
\end{equation} where the integration is over along wires within a unit cell (abbreviated as ``$uc$''). Unit cell ``area'' are total lengths within a unit cell which equals $2a_x$.

For a stripe S, an vertial array of unit cells that host a single moir\'e wire (1d), Fourier transformed interaction potential at momentum $q$

\begin{equation}
\tilde{V}(q)=a_y \sum_{\alpha,\alpha'=\pm}\int_{uc} d\mathbf{r} \int_S d\mathbf{r}'  \psi_{\alpha,q_1+q}^\dag(\mathbf{r})\psi_{\alpha,q_1}(\mathbf{r})\psi_{\alpha',q_2-q}^\dag(\mathbf{r}')\psi_{\alpha',q_2}(\mathbf{r}') V(\mathbf{r}-\mathbf{r}') 
\end{equation} where $q_x$ is omitted for simplicity. 

With 
\begin{equation}
\psi_{\alpha,q_y}(\mathbf{r}+a_y \hat{y})=e^{i q_y a_y} \psi_{\alpha,q_y}(\mathbf{r}),
\end{equation}  and if there are further $n$th neighbour interactions in $y$ direction with $V_I^n$ is symmetric in $y$-direction then the two Fourier components of potential at $k=0$ and $2q_F$ are

\begin{equation}
\begin{aligned}
&\widetilde{V}(0)=a_y\sum_{\alpha,\alpha'=\pm} \int_{uc} d\mathbf{r} \int_{uc} d\mathbf{r}'  \psi_{\alpha,q_F}^\dag(\mathbf{r})\psi_{\alpha,,q_F}(\mathbf{r})\psi_{\alpha',-q_{F}}^\dag(\mathbf{r}')\psi_{\alpha',-q_{F}}(\mathbf{r}') (\sum_{n=-\infty}^\infty V_I(\mathbf{r}-\mathbf{r}'+n a_y \hat{y}))\\
&\widetilde{V}(2q_F)=a_y\sum_{\alpha,\alpha'=\pm} \int_{uc} d\mathbf{r} \int_{uc} d\mathbf{r}' \psi_{\alpha,q_F}^\dag(\mathbf{r})\psi_{\alpha,-q_{F}}(\mathbf{r})\psi_{\alpha',-q_{F}}^\dag(\mathbf{r}')\psi_{\alpha',q_{F}}(\mathbf{r}') (\sum_{n=-\infty}^\infty V_I(\mathbf{r}-\mathbf{r}'+n a_y \hat{y})\cos{(2nq_Fa_y)})
\end{aligned}
\end{equation}
where 
\begin{equation}
V_I(\br)=U \exp\left(-\sqrt{\frac{x^2}{x_0^2}+\frac{y^2}{y_0^2}+\frac{z^2}{z_0^2}}\right)
\end{equation}
as we defined in the main text, where $x_0=1$nm, $y_0=5$nm, $z_0=10$nm. For our numerical calculations, we take a cutoff of $|n|\le 6$ nearest unit cells along $y$-direction.

\section{Monolayer $WTe_2$ Luttinger parameter estimation}
\label{AppC}
In this section, we give a brief estimation of Luttinger parameter for a single layer of $WTe_2$ following similar reasons we give in this paper. 

In monolayer case, we have our interaction potential along the direction of the wire ($x$ direction) in momentum space as

\begin{equation}
V_I(k)=\frac{2 U x_0}{1+k^2x_0^2}
\end{equation} where $x_0=1nm$.
Therefore, the effective Luttinger parameter is again given by
\begin{equation}
g_{\text{eff}}=\sqrt{\frac{2\pi v_F^{(0)}+V_I(2k_F)}{2\pi v_F^{(0)}+2V_I(0)-V_I(2k_F)}},
\end{equation} 
where $v_F^{(0)}=\frac{\hbar^2 k_F}{m}$ is the Fermi velocity of the microscopic wire, 
and $\hbar k_F=\sqrt{2mE_F}$. For different interaction strength $U$ with $E_F=30meV$, we have 
\begin{center}
\begin{tabular}{ |c|c|c|c|c|c|c| } 
\hline
 & $U=0.2eV$ & $0.5eV$ & $1eV$ & $2eV$ & $3eV$ & $4eV$   \\
\hline
$g_{\text{eff}}$ & 0.82 &  0.72 & 0.65 & 0.6 &0.59 & 0.58\\ 

\hline
\end{tabular}
\end{center}

\section{Interwire hopping parameter estimation from ab initio calculations}
\label{AppD}
The first-principle calculations were performed on the Vienna ab initio simulation package\cite{vasp1,vasp2}. The generalized gradient approximation with the Perdew-Burke-Ernzerhof type exchange-correlation potential was adopted\cite{PBE}. The convergence accuracy of self-consistent calculations is $10^{-6}$ eV per unit cell by using $k$ grids with a $11\times11\times1$ mesh. We constructed the tight-binding Hamiltonian of the monolayer WTe$_2$ using the Wannier90 package\cite{wannier90} and generated the maximally localized Wannier functions for $5d$ orbitals on tungsten and $5p$ orbitals on tellurium.

Based on the tight-binding Hamiltonian from Wannier calculations, we estimate the interwire hopping parameter $t_\perp$ within each layer. 
Under the Wannier function basis, i.e., $\{d^{\alpha}_{z^2}, d^{\alpha}_{xz}, d^{\alpha}_{yz}, d^{\alpha}_{x^2-y^2}, d^{\alpha}_{xy}\}$ where $\alpha$ represents the two W atoms in one unit cell, the wave function of the top valence band at the $\Gamma$ point $\ket{\Psi}_v$ is expressed as,
\begin{equation}
    \ket{\Psi}_v = \sum_i c_{i} \ket{\phi}_i
\end{equation}
where $i=1,2,...,10$ is the index of Wannier function basis $\ket{\phi}_i$.  
Then the interwire hopping parameter $t_\perp$ is estimated by the following equation,
\begin{equation}
    t_\perp = \sum_{m \in C_1, n \in C_2} c_m^{\star} c_n t_{mn}
\end{equation}
where $t_{mn}$ is the hopping strength between the Wannier orbital $\ket{\psi_m}$ in wire $C_1$ and the Wannier orbital $\ket{\psi_n}$ in wire $C_2$. This yields $t_\perp\approx 20$meV.

\section{Interlayer hopping of strength}
\label{AppE}
\begin{figure}
    \centering    \includegraphics{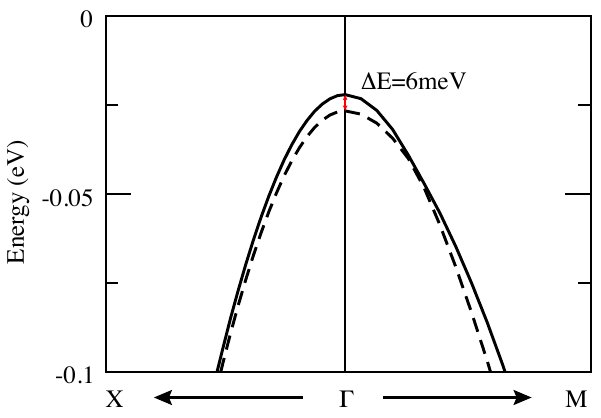}
    \caption{Band structure calculation for a bilayer-structure WTe$_2$ nearby the $\Gamma$ point. As indicated by the red arrow line, due to the interlayer coupling, there is a hybridization band gap $\Delta E = 6$ meV for the top valence bands around the $\Gamma$ point.}
    \label{fig:bilayerband}
\end{figure}

Based on the first principle calculations, we also obtained the interlayer hopping of strength. As shown in Fig. \ref{fig:bilayerband}, the hybridization band gap between the top and bottom WTe$_2$ layers is about $\Delta E = 6$ meV. In the effective moir\'e model we consider, the delta function interlayer hopping strength $\lambda'$ is roughly the interlayer hopping energy times the spatial range of the interlayer hopping. For two wires in different layers, the two wires are close within the length scale of the $x$ direction moir\'e lattice constant $a_x=\frac{l}{2\sin(\theta/2)}$, so we regard $a_x$ as the spatial range of interlayer hopping. For angles $\theta\lesssim 5^\circ$ we consider, with $l=0.627$nm, we have $a_x\gtrsim6$nm, and thus we estimate the delta function interlayer hopping strength as $\lambda'\approx \Delta E a_x/2 \approx 20 \text{ meV} \cdot \text{nm}$. 

For the intralayer delta function potential $\lambda$ from the other layer, we adopt $\lambda\approx 10 \text{ meV} \cdot \text{nm}$, which has the same order of magnitude magnitude following a similar argument as above.

We note that these parameters are by no means aimed to characterize WTe$_2$ accurately, since the coupled wire model we study is an oversimplified description to such moir\'e systems. The procedure is to ensure the parameters are within the reasonable ranges of physical materials.

\twocolumngrid

\end{document}